\documentclass[12pt]{article}

\usepackage{amsmath}
\usepackage{graphicx}
\usepackage{amsfonts}
\usepackage{amssymb}
\usepackage{latexsym}
\usepackage{color}
\usepackage{cite}
\input{colordvi.tex}

\renewcommand{\thefootnote}{\#\arabic{footnote}}

\begin{document}
\setcounter{footnote}{0}

\begin{titlepage}
\begin{flushright}
RESCEU-43/12
\end{flushright}
\begin{center}


\vskip .5in

{\Large \bf
Consequences of a stochastic approach to the conformal invariance of inflationary correlators
}
\vskip .45in

{\large
Hayato Motohashi$^{1,2}$,
Teruaki Suyama$^2$
and 
Jun'ichi Yokoyama$^{2,3}$
}

\vskip .45in%

{\em 
$^1$
  Department of Physics, Graduate School of Science,
  The University of Tokyo, Tokyo 113-0033, Japan
  }\\
{\em
$^2$
  Research Center for the Early Universe (RESCEU), Graduate School
  of Science,\\ The University of Tokyo, Tokyo 113-0033, Japan
  }\\
{\em
$^3$
  Kavli Institute for the Physics and Mathematics of the Universe (Kavli IPMU),\\
  The University of Tokyo, Kashiwa, Chiba, 277-8568, Japan
  }

\end{center}

\vskip .4in

\begin{abstract}
We provide a general formalism to calculate the infrared correlators of multiple interacting scalar fields
in the de Sitter space by means of the stochastic approach.
These scalar fields are treated as test fields and hence our result is applicable to 
the models such as the curvaton scenario where the fields that yield initially isocurvature modes
do not contribute to the cosmic energy density during inflationary expansion.
The stochastic formalism combined with the argument of conformal invariance of the correlators 
reflecting the de Sitter isometries
allows us to fix the form and amplitude of the three-point functions completely and partially for
the four-point functions in terms of calculable quantities.
It turns out that naive scaling argument employed in the previous literature does not necessarily hold
and we derive the necessary and sufficient condition for the correlator to
obey the naive scaling.  
We also find that correlation functions can in principle exhibit more complicated 
structure than argued in the literature.
\end{abstract}
\end{titlepage}

\renewcommand{\thepage}{\arabic{page}}
\setcounter{page}{1}
\renewcommand{\thefootnote}{\#\arabic{footnote}}

\section{Introduction}
De Sitter spacetime is as fundamental as Minkowski spacetime.
It describes the accelerated expansion of the Universe sourced by the cosmological constant,
or the effective vacuum energy mimicked by some scalar field.
In addition to the observation that present Universe is undergoing accelerated expansion,
it has become a standard paradigm that the small inhomogeneity, 
{\it i.e.}, tiny deviation from the Friedmann-Lema\^itre-Robertson-Walker
Universe, is also thought to be explained by conversion of the quantum fluctuations of 
light scalar fields generated during de Sitter expansion in the early Universe \cite{Liddle-Lyth}.
Because of this, it is important to study statistical properties of the scalar
field in de Sitter space.

The simplest scenario for the generation of the curvature perturbation is to assume that 
inflaton which causes inflation simultaneously generates observed amplitude of the primordial 
curvature perturbation $\sim 10^{-5}$ \cite{Liddle-Lyth}.
Despite its simplicity and that its predictions are consistent with observations \cite{Komatsu:2010fb}, 
this scenario is not a prediction of (unknown) fundamental theory but rather an assumption.
Actually, there are also many alternatives to this scenario most of which introduce other scalar
fields which are originally isocurvature modes and convert to the curvature perturbation
after inflation.
The curvaton \cite{Enqvist:2001zp,Lyth:2001nq,Moroi:2001ct} 
and the modulated reheating models \cite{Dvali:2003em,Kofman:2003nx} 
are the representative ones that
belong to this category.
The scalar fields in this category are generically negligible during inflation, that is, 
they do not affect the dynamics of the inflationary expansion.
Because of this, the correlation functions among such scalar fields become invariant 
under the de Sitter isometries.
Note that this is not necessarily true for the case of the inflaton fluctuation since the inflaton
can affect the expansion of the Universe.

The purpose of this paper is to study generic statistical properties of the correlation
functions of scalar fields which enjoy full de Sitter invariance.
In particular, we are concerned with infrared limit of the correlators.
This is equivalent to large distance limit or to late time limit in de Sitter space
since any two points that are close originally is eventually stretched to arbitrary
large distance by accelerated expansion.

Some literature \cite{Antoniadis:1996dj,Antoniadis:2011ib,Creminelli:2011mw,Kehagias:2012pd} 
already discuss the generic shape of the correlators consistent with de Sitter invariance. 
In these papers, the scalar fields appearing in the correlators are (implicitly) assumed to scale
like $(-\eta)^{\Delta_a}$ for $-\eta \ll H^{-1}$ (late time),
where $\eta$ is the conformal time of the de Sitter metric given by
\begin{equation}
ds^2=\frac{-d\eta^2+d{\vec x}^2}{H^2 \eta^2},
\end{equation}
and $\Delta_a$ is a constant specific to the scalar field $\phi_a$ and 
cannot be constrained only by the de Sitter invariance.
For instance, $\langle \phi_a (\eta_1,{\vec x_1}) \phi_b (\eta_2,{\vec x_2}) \rangle \sim {(-\eta_1)}^{\Delta_a} {(-\eta_2)}^{\Delta_b}$ etc., 
where $\Delta_a$ neither depends on the number of fields nor the type of fields appearing
in the correlator under consideration but just depends on $\phi_a$.
This time dependence of the correlators then allows us to (partially) fix their dependence on
the spatial coordinates by requiring that the correlators are invariant under the
de Sitter isometric transformations.
However, it is not clear how wide class of models satisfy the above scaling.

In order to evaluate the correlation functions for the de Sitter invariant state
without introducing the scaling assumption a priori,
we make use of the stochastic formalism.
This formalism was introduced and developed in \cite{Starobinsky:1982ee,Starobinsky:1984fx,Starobinsky:1986fx,Sasaki:1987gy,Starobinsky:1994bd}.
It has been since then employed for various inflationary models to study the 
infrared behavior of the scalar fields ({\it e.g.},\cite{Nakao:1988yi,Nambu:1988je,Mollerach:1990zf,Starobinsky:1994bd,Martin:2005ir,Hattori:2005ac,Finelli:2010sh,Lorenz:2010vf,Martin:2011ib,Kawasaki:2012bk}).
The formalism has been also used to determine the distribution of the initial value of 
the scalar field that may become important in the late Universe.
This is important, for example, to provide the initial value of the quintessence field
which causes the accelerated expansion of the present Universe \cite{Martin:2004ba,Ringeval:2010hf}.

The stochastic formalism solves the infrared dynamics by treating the long wavelength modes
as the classical statistical variables that are sourced by stochastic noises coming from the
short wavelength quantum modes.
Since the dynamics can be solved, we can fix the form of the correlation functions unlike
in the literature where they are constrained only by the argument of the de Sitter invariance.
The stochastic formalism is especially useful when the nonperturbative effect is crucial
to obtain the correct correlation functions as in the case of massless self-interacting 
scalar field \cite{Starobinsky:1994bd}.
This is simply because infrared dynamics is solved without invoking perturbative 
expansion, that is, nonperturbative effect is automatically taken into account
in the stochastic formalism.

Despite the many existing applications of the stochastic formalism to particular inflationary models,
we do not find any paper that discusses general consequences of the stochastic approach on
the infrared properties of the correlators of the multiple interacting scalar fields,
which motivated us to address this issue.
This article is the report of the calculations of the two-, three- and four-point functions
of the multiple scalar fields derived by the use of the stochastic formalism.
As will be demonstrated explicitly,
in addition to the scaling index of the correlators,
their amplitudes, 
which are completely unconstrained within the framework of the symmetry argument,
can also be expressed in terms of quantities that are reasonably calculable
in the stochastic approach.
Thus we can completely fix the (infrared) correlation functions.
We will find that the naive universal scaling $\phi_a \sim (-\eta)^{\Delta_a}$
in the correlators does not always hold and provide the necessary condition for
this scaling to hold by explicitly constructing three-point and four-point functions.

\section{Basics of the stochastic formalism}
Our purpose is to study superhorizon evolution of (weakly) interacting multiple fields in de Sitter space
and to calculate the resulting correlation functions in the stochastic formalism.
Before doing this, in this section, let us briefly review the basic points of
this formalism.
For more details, see, for example, \cite{Starobinsky:1982ee,Starobinsky:1984fx,Starobinsky:1986fx,Sasaki:1987gy,Starobinsky:1994bd}.

The first step of the stochastic formalism is to split the field (operator) into
the long wavelength part ${\pmb \phi}_L$ and the short wavelength part;
\begin{equation}
{\pmb \phi} (t,{\vec x})={\pmb \phi}_L (t,{\vec x})+\int \frac{d^3 k}{{(2\pi)}^{\frac{3}{2}}} \theta (k-\epsilon a(t) H) \left( {\pmb a}_{\vec k} {\pmb \phi}_{\vec k}(t)e^{-i {\vec k} \cdot {\vec x}}+h.c. \right),
\end{equation}
where $\theta (x)$ is a step function and
we have used a bold letter ${\pmb \phi} =(\phi_1,\phi_2,\cdots,\phi_N)$ 
to make it refer to the multiple scalar fields.
The second term on the right-hand side represents the contribution from short
wavelength modes whose wavenumbers are greater than $\epsilon a(t) H$.
Since we are assuming that the background spacetime is de Sitter one ($a(t) = e^{Ht}$),
any mode that originally belongs to the short wavelength part eventually enters ${\pmb \phi}_L$.
Although the evolution of the short wavelength part depends on the nature of the interactions,
in this paper, we assume that any interaction is negligible for the short wavelength part
and ${\pmb \phi}_{\vec k}$ obeys the massless Klein-Gordon equation in de Sitter space whose solution is given by
\begin{equation}
\phi_{a,{\vec k}}=\frac{H}{\sqrt{2k}} \left( \eta-\frac{i}{k} \right) e^{-ik\eta},
\end{equation}
where the subscript $a$ of $\phi$ runs from $1$ to $N$ and $\eta =-\frac{1}{aH}$ is the conformal time.
Then we can treat ${\pmb a}_{\vec k}$ and ${\pmb a}^\dagger_{\vec k}$ as the standard
annihilation and creation operators.

The second step is to consider the evolution equation for the long wavelength part ${\pmb \phi}_L$.
Applying the slow-roll approximation and neglecting the higher-order terms in the
short-wave modes and mode mixing terms, it reads
\begin{equation}
{\dot \phi_a}(t,{\vec x}) = -\frac{1}{3H}V_a({\pmb \phi})+f_a(t,{\vec x}), \label{eom-phi}
\end{equation}
where $V_a\equiv \partial V/\partial \phi_a$,
and we have abbreviated the subscript $L$ for the coarse-grained field.
This equation may be regarded as a classical Langevin equation with a stochastic noise term,
$f_a$, which is given by time derivative of the short wavelength part and
represents modes of $k=\epsilon a(t) H$ that join ${\pmb \phi}_L$ at time $t$. 
Straightforward calculation shows that $f_a$ is random Gaussian whose two-point 
function is given by
\begin{equation}
\langle f_a (t_1,{\vec x_1}) f_b (t_2,{\vec x_2}) \rangle =\delta_{ab} \frac{H^2}{4\pi^2} \delta (t_1-t_2) j_0 (\epsilon a(t) H |{\vec x_1}-{\vec x_2}|),
\end{equation}
where $j_0(x) \equiv \frac{\sin x}{x}$ is the unnormalized sinc function.

Then we can easily show that the one-point probability density 
$\rho_1 ({\pmb \phi}({\vec x}),t)\equiv \rho (\phi_1({\vec x}),\phi_2({\vec x}),\cdots,t)$ 
obeys the Focker-Planck equation:
\begin{equation}
\frac{\partial}{\partial t} \rho_1 ({\pmb \phi},t) =\frac{\partial}{\partial \phi_a} \left( \frac{V_a}{3H} \rho_1 ({\pmb \phi},t) \right)+\frac{H^3}{8\pi^2} \delta_{ab} \frac{\partial^2}{\partial \phi_a \partial \phi_b} \rho_1 ({\pmb \phi},t).
\end{equation}
Introducing the dimensionless potential by $v({\pmb \phi}) \equiv 4\pi^2 V({\pmb \phi})/(3H^4)$,
a general solution of the above equation can be written as 
\begin{equation}
\rho_1 ({\pmb \phi},t)=e^{-v({\pmb \phi})} \sum_n a_n \Phi_n ({\pmb \phi}) e^{-\Lambda_n (t-t_0)}, \label{sol-rho1}
\end{equation}
where $\Phi_n ({\pmb \phi})$ is an eigenfunction of $N$-dimensional Schr\"odinger equation:
\begin{equation}
\sum_a \left( -\frac{1}{2} \frac{\partial^2}{\partial \phi_a \partial \phi_a}+\frac{1}{2} (v_a v_a -v_{aa}) \right) \Phi_n ({\pmb \phi})=\frac{4\pi^2 \Lambda_n}{H^3} \Phi_n ({\pmb \phi}). \label{Schrodinger}
\end{equation}
To be definite, we only consider the case the eigenfunctions are normalizable and satisfy 
\begin{equation}
\int d{\pmb \phi} ~\Phi_m ({\pmb \phi})\Phi_n ({\pmb \phi})=\delta_{mn}.
\end{equation}
The left-hand side of Eq.~(\ref{Schrodinger}) can be written as
\begin{equation}
\sum_a \frac{1}{2} \left( -\frac{\partial}{\partial \phi_a}+v_a \right) \left( \frac{\partial}{\partial \phi_a}+v_a \right) \Phi_n ({\pmb \phi}).
\end{equation}
By multiplying $\Phi_n ({\pmb \phi})$ to the above expression from left and integrating it by parts, the left-hand side of \eqref{Schrodinger} becomes the integral of $\left[\left( \frac{\partial}{\partial \phi_a}+v_a \right) \Phi_n ({\pmb \phi})\right]^2$.
Therefore, $\Lambda_n \ge 0$. 
In particular, the eigenfunction $\Phi_0$ having the minimum eigenvalue ({\it i.e.}, $\Lambda_0=0$)
is given by
\begin{equation}
\Phi_0 ({\pmb \phi})={\cal N} e^{-v({\pmb \phi})},~~~~~{\cal N}={\left( \int d{\pmb \phi}~e^{-2v({\pmb \phi})} \right)}^{-\frac{1}{2}}.
\end{equation}
At sufficiently late time, all the modes having the positive $\Lambda_n$ decays in
Eq.~(\ref{sol-rho1}) and $\rho_1$ becomes independent of time;
\begin{equation}
\rho_1 ({\pmb \phi},t) \to \rho_{\rm eq} ({\pmb \phi})={\cal N}^2 e^{-2 v({\pmb \phi})}. \label{de-S-inv}
\end{equation}
Here we used $a_0={\cal N}$ which holds from the normalization $\int d{\pmb \phi} \rho_1({\pmb \phi},t) =1$.
This is a distribution function for an equilibrium state achieved in the de Sitter space.

It is possible that the integral appearing in the definition of ${\cal N}$ diverges.
In such a case, Eq.~(\ref{sol-rho1}) does not possess any static solution and hence there
is no equilibrium state.
This happens, for example, for massless free scalar field.

\section{Correlation functions}
The formalism explained above enables us to evaluate correlation functions between multiple fields.
In the following, we will derive the infrared behaviors of the two-point, three-point and four-point
functions separately by means of the stochastic formalism.
Derivation of the expression of the two-point functions mostly follows 
the one developed in \cite{Starobinsky:1994bd},
which also showed their de Sitter invariance constructed from the equilibrium distribution function.

\subsection{Two-point functions}
The (de Sitter invariant) spatial correlators (on superhorizon scales) at equal time can be written as
\begin{equation}
\langle \phi_a (t,{\vec x_1}) \phi_b (t,{\vec x_2}) \rangle=\int d{\pmb \phi}^1 
d{\pmb \phi}^2~\phi^1_a \phi^2_b ~\rho_2({\pmb \phi}^1,{\pmb \phi}^2,t), \label{two-point0}
\end{equation}
where $\rho_2({\pmb \phi}^1,{\pmb \phi}^2,t)$ is the probability
density of finding ${\pmb \phi}^i$ at ${\vec x_i}$ ($i=1,2$).
It can be shown that $\rho_2$ obeys the following equation,
\begin{eqnarray}
\frac{\partial \rho_2}{\partial t}=\sum_{i=1}^2 \left[ \frac{\partial}{\partial \phi_a^i} \left( \frac{V_a ({\pmb \phi}^i)}{3H} \rho_3 \right)+\frac{H^3}{8\pi^2} \frac{\partial^2 \rho_3}{\partial \phi_a^i \partial \phi_a^i} \right]+\frac{H^3}{4\pi^2} \frac{\partial^2 \rho_3}{\partial \phi_a^1 \partial \phi_a^2} j_0 (\epsilon a(t) H|{\vec x_1}-{\vec x_2}|). \label{evo-rho2}
\end{eqnarray}
At early times when the points ${\vec x_1}$ and ${\vec x_2}$ are deeply inside 
the same Hubble patch,
$j_0$ on the right-hand side becomes unity.
For such a case, it can be shown that 
\begin{equation}
\rho_2 ({\pmb \phi}^1,{\pmb \phi}^2,t)= \delta ({\pmb \phi}^2-{\pmb \phi}^1) \rho_{\rm eq}({\pmb \phi}^1), \label{two-eq}
\end{equation}
constitutes a static solution.
This is very reasonable since setting $j_0=1$ means that ${\pmb \phi_1}$ and
${\pmb \phi_2}$ are fully correlated, {\it i.e.}, ${\pmb \phi_1}={\pmb \phi_2}$
and the appearance of $\rho_{\rm eq}$ reflects Eq.~(\ref{de-S-inv}) that guarantees the de Sitter invariance \cite{Starobinsky:1994bd}.
Therefore, Eq.~(\ref{two-eq}) can be used as the initial condition for Eq.~(\ref{evo-rho2}).

It is hard to find the analytic solution of Eq.~(\ref{evo-rho2}) with the initial
condition (\ref{two-eq}).
Here we make an approximation that
\begin{equation}
j_0 (\epsilon a(t) Hr)\simeq \theta (1-\epsilon a(t) Hr).
\end{equation}
This drastically simplifies the equation without losing any essential
point of the stochastic formalism.
This approximation allows us to write down the solution of Eq.~(\ref{evo-rho2})
after ${\vec x_1}$ and ${\vec x_2}$ are separated by super-horizon distance;
\begin{equation}
\rho_2 ({\pmb \phi}^1,{\pmb \phi}^2,t)=
\int d{\pmb \phi}_r \Pi ({\pmb \phi}^1,t;{\pmb \phi}_r,t_r) 
\Pi ({\pmb \phi}^2,t;{\pmb \phi}_r,t_r) \rho_{\rm eq} ({\pmb \phi}_r),
\end{equation}
where $t_r$ is a solution of $\epsilon a(t_r) H|{\vec x_1}-{\vec x_2}|=1$ and
represents the time when ${\pmb \phi}$ at ${\vec x_1}$ and ${\pmb \phi}$ at ${\vec x_2}$
get uncorrelated.
Here $\Pi ({\pmb \phi}^1,t_1;{\pmb \phi}^2,t_2)$ is the transition probability
from ${\pmb \phi}={\pmb \phi}^2$ at $t=t_2$ to ${\pmb \phi}={\pmb \phi}^1$ at $t=t_1$.
Its expression in terms of the eigenfunctions is given by \cite{Starobinsky:1994bd}
\begin{equation}
\Pi ({\pmb \phi}^1,t_1;{\pmb \phi}^2,t_2)=e^{-v({\pmb \phi}^1)+v({\pmb \phi}^2)} \sum_{n=0}^\infty \Phi_n ({\pmb \phi}^1) \Phi_n ({\pmb \phi}^2) e^{-\Lambda_n (t_1-t_2)}. \label{sol-Pi}
\end{equation}
This expression of $\rho_2$ appeals to our intuition,
that is, $\rho_2$ is given by the product of the probability
of ${\pmb \phi}$ going to ${\pmb \phi_1}$ from ${\pmb \phi_r}$ by
the stochastic process described by Eq.~(\ref{eom-phi})
and that of ${\pmb \phi}$ going to ${\pmb \phi_2}$ from ${\pmb \phi_r}$
with a weight $\rho_{\rm eq}({\pmb \phi_r})$.
Using this picture for $\rho_2$, we have
\begin{eqnarray}
\langle \phi_a (t,{\vec x_1}) \phi_b (t,{\vec x_2}) \rangle = \int  d{\pmb \phi}^1 d{\pmb \phi}^2 ~\phi^1_a \phi^2_b \int d{\pmb \phi}_r \Pi ({\pmb \phi}^1,t;{\pmb \phi}_r,t_r) \Pi ({\pmb \phi}^2,t;{\pmb \phi}_r,t_r) \rho_{\rm eq} ({\pmb \phi}_r). \label{two-point}
\end{eqnarray}
Substituting Eq.~(\ref{sol-Pi}) into Eq.~(\ref{two-point}), we find
\begin{equation}
\langle \phi_a (t,{\vec x_1}) \phi_b (t,{\vec x_2}) \rangle ={\cal N}^2 \sum_{n=0}^\infty A^{(n)}_{a} A^{(n)}_{b} {(HR_{12})}^{-\frac{2 \Lambda_n}{H}} \exp \left( -\frac{2\Lambda_n}{H} \ln \epsilon \right), \label{two-point-2}
\end{equation}
where $R_{12}=a(t)|{\vec x_1}-{\vec x_2}|$ is the physical distance
between ${\vec x_1}$ and ${\vec x_2}$ and $A^{(n)}_a$ is defined by
\begin{equation}
A^{(n)}_a = \int d {\pmb \phi}~\phi_a e^{-v({\pmb \phi})} \Phi_n ({\pmb \phi}). \label{def-A}
\end{equation}
Equation (\ref{two-point-2}) is our expression for the two-point functions.
To minimize the effect of $\epsilon$, 
we choose $\epsilon$ so that it satisfies $\exp \left( -\frac{2\Lambda_n}{H} \ln \epsilon \right) \sim 1$
for the dominant mode contributing to the two-point functions,
as suggested in \cite{Starobinsky:1994bd}.
Now let us consider correlation function (\ref{two-point-2}) on sufficiently late time or
(equivalently) large scales in which case $a(t) |{\vec x_1}-{\vec x_2}|/H^{-1}$ is quite large.
Then the leading contribution to the correlator is from a state $\Phi_{\bar n}({\pmb \phi})$ labeled
by an integer ${\bar n}$ having minimum 
$\Lambda_{\bar n}$ (apart from the ground state $n=0$ which has $\Lambda_n=0$) with nonvanishing $A^{({\bar n})}_{a} A^{({\bar n})}_{b}$.
Generally, ${\bar n}$ depends on the choice of fields $(\phi_a,~\phi_b)$ and can vary for different set of fields of the correlators.
In particular, 
it may happen that ${\bar n}$ for $A^{({\bar n})}_{a} A^{({\bar n})}_{a}$ which
we denote by 
$n_a$ is different from $n_b$ and $\Lambda_{n_a} \neq \Lambda_{n_b}$. 
In such a case, ${\bar n}$ for $A^{({\bar n})}_{a} A^{({\bar n})}_{b}$ which we denote by 
$n_{ab}$
may or may not exist.
If $n_{ab}$ does not exist, it means no correlation between $\phi_a$ and $\phi_b$.
Alternatively, our result can also allow a possibility that $n_{ab}$ exists,
in which case $\phi_a$ and $\phi_b$ are correlated.
By a simple consideration, we find that such $n_{ab}$ is either equal to or larger than $n_a$ or $n_b$ whichever is greater.
In short, two different fields having different ${\bar n}$ ({\it i.e.}, $n_a \neq n_b$)
can in principle have correlation between them,
which is consistent with our findings for the simple example demonstrated
in the Introduction.

On the contrary, 
if some of the fields are interacting so that the minimum integers for the nonvanishing of $A^{(n)}_a$ 
for such fields are all the same,
then the correlator exhibits a universal behavior in the sense that
the scaling index for any field takes the same value and
is completely given by $\Lambda_{\bar n}$, eigenvalue of the $N$-dimensional
Schr\"odinger equation.
In this case, correlators at late time scales as
\begin{equation}
\langle \phi_a (t,{\vec x_1}) \phi_b (t,{\vec x_2}) \rangle \sim {|{\vec x_1}-{\vec x_2}|}^{-\frac{2\Lambda_{\bar n}}{H}}.
\end{equation}

\subsection{Three-point functions}
What we want to evaluate is the spatial three-point functions evaluated
at equal time $t$;
\begin{equation}
\langle \phi_a (t,{\vec x_1}) \phi_b (t,{\vec x_2}) \phi_c (t,{\vec x_3}) \rangle=\int d{\pmb \phi}^1 d{\pmb \phi}^2 d{\pmb \phi}^3~\phi^1_a \phi^2_b \phi^3_c ~\rho_3 ({\pmb \phi}^1,{\pmb \phi}^2,{\pmb \phi}^3,t), \label{three-point}
\end{equation}
where $\rho_3({\pmb \phi}^1,{\pmb \phi}^2,{\pmb \phi}^3,t)$ is the probability
density of finding ${\pmb \phi}^i$ at ${\vec x_i}$ ($i=1,2,3$).
It can be shown that $\rho_3$ obeys the following equation,
\begin{eqnarray}
\frac{\partial \rho_3}{\partial t}=\sum_{i=1}^3 \left[ \frac{\partial}{\partial \phi_a^i} \left( \frac{V_a ({\pmb \phi}^i)}{3H} \rho_3 \right)+\frac{H^3}{8\pi^2} \frac{\partial^2 \rho_3}{\partial \phi_a^i \partial \phi_a^i} \right]+\frac{H^3}{4\pi^2} \sum_{i<j} \frac{\partial^2 \rho_3}{\partial \phi_a^i \partial \phi_a^j} j_0 (\epsilon a(t) H|{\vec x_i}-{\vec x_j}|). \label{evo-rho3}
\end{eqnarray}
At early times, all the three points are inside the Hubble radius,
{\it i.e.}, $a(t)|{\vec x_i}-{\vec x_j}|H \ll 1$ for any $1 \le i,j \le 3$, and
the fields are maximally correlated each other.
During this epoch, the evolution equation for $\rho_3$ can therefore be well approximated
by Eq.~(\ref{evo-rho3}) with all the $j_0$ being replaced by unity.
We can verify that this equation allows the following solution,
\begin{equation}
\rho_3 ({\pmb \phi}^1,{\pmb \phi}^2,{\pmb \phi}^3,t)=\delta ({\pmb \phi}^3-{\pmb \phi}^2) \delta ({\pmb \phi}^2-{\pmb \phi}^1) \rho_{\rm eq}({\pmb \phi}^1),
\end{equation}
whose physical meaning is obvious from the reasoning we made earlier.
This solution is independent of time and can be used as a de Sitter invariant 
initial condition of Eq.~(\ref{evo-rho3}).
Then the problem is to solve Eq.~(\ref{evo-rho3}) with such an initial condition
until sufficiently late time when all the points are separated by super-horizon length 
and any correlation between
the fields at different points is turned off.
Although this is a well defined mathematical problem and we can in principle
solve Eq.~(\ref{evo-rho3}) and perform the integrals appearing in Eq.~(\ref{three-point})
to get the three-point functions,
we find it difficult in practice to solve Eq.~(\ref{evo-rho3}) which is highly
involved partial differential equation.
Fortunately, as long as we are only concerned with sufficiently late
time behavior of the three-point functions,
which is actually the present case,
there is a way to derive an analytic expression without directly solving
Eq.~(\ref{evo-rho3}) as we will demonstrate below.
The point is to utilize the de Sitter isometries which allows us to find 
the three-point functions at general points by implementing the coordinate
transformation that preserves the de Sitter metric from some extreme 
configuration of the points where analytic evaluation of the three-point
functions (to a very good approximation) is feasible.

As is well known, there are 10 isometries for the metric of the de Sitter
spacetime.
Among 10 isometries, translations and rotations for the spatial coordinate
constitute 6 isometries.
We also have the dilatation isometry which amounts to multiply both
time and spatial coordinates by the same constant factor.
The remaining 3 isometries are complex mixing whose infinitesimal form is given by
\begin{equation}
\eta'=\eta-2\eta ({\vec b} \cdot {\vec x}),~~~~~{\vec x}'={\vec x}+{\vec b}(-\eta^2+x^2)-2 ({\vec b} \cdot {\vec x}) {\vec x},
\end{equation}
where $x^2 \equiv{\vec x}^2$ and ${\vec b}$ is an infinitesimal constant vector.
On super-horizon scales, or on sufficiently late time, in which 
$\eta$ is much smaller than ${\vec x}$,
the finite version of the above transformation can be written as
\begin{equation}
\eta'=\frac{\eta}{1+2{\vec b}\cdot {\vec x}+b^2 x^2},~~~~~{\vec x}'=\frac{{\vec x}+x^2 {\vec b}}{1+2{\vec b}\cdot {\vec x}+b^2 x^2}. \label{transform}
\end{equation}
The transformation of the spatial coordinate does not involve time
and becomes exactly what is known as the special conformal transformation.
The special conformal transformation combined with the dilatation,
rotation and translation transformations constitute the conformal
transformation \cite{francesco}.
We will come back to this point later when we utilize the
conformal symmetry to fix the correlators.

Now let us consider the three-point function in the squeezed limit with 
different time coordinates for the different points;
\begin{equation}
\langle \phi_a (t_1,{\vec y_1}) \phi_b (t_2,{\vec y_2}) \phi_b (t_3,{\vec y_3}) \rangle. \label{3pt-squeezed}
\end{equation}
Although what we are interested in is the equal time correlators ($t_1=t_2=t_3=t$),
for the moment, we let them to be independent due to the reason which will become
clear later.
To be definite, we take ${\vec y_3}={\vec 0}$ and $|{\vec y_1}| \gg |{\vec y_2}|$ (squeezed limit),
which is always possible without a loss of generality.
For convenience, we define $R\equiv |{\vec y_1}|$ and $r\equiv |{\vec y_2}|$. 
All the time coordinates $t_1,t_2$ and $t_3$ are assumed to be very large so that
any two different points are eventually separated by super-horizon size distance. 
Instead of directly solving Eq.~(\ref{evo-rho3}), 
the following physical consideration enables us to evaluate Eq.~(\ref{3pt-squeezed}).
By definition, ${\vec y_2}$ and ${\vec y_3}$(this is actually ${\vec 0}$) 
are close together compared to ${\vec y_1}$.
Therefore, a field at ${\vec y_1}$ first gets uncorrelated and starts to
evolve independently when the physical distance between ${\vec y_1}$ and other two points becomes
equal to ${(\epsilon H)}^{-1}$.
Strictly speaking, the epoch when $a H |{\vec y_1}-{\vec y_2}|=1$ occurs is different
from that when $a H |{\vec y_1}-{\vec y_3}|=1$ is satisfied. 
But the difference between these little affects the final result in the squeezed limit
and we can safely take them as being equal.
At this time, ${\vec y_2}$ and ${\vec y_3}$ are still deeply inside the Hubble radius
and fields at those two points take the same value.
As the Universe expands, the physical distance between ${\vec y_2}$ and ${\vec y_3}$
then becomes equal to ${(\epsilon H)}^{-1}$.
After this time, all the fields at different points evolve separately.
This picture, which is a good approximation when there is a huge hierarchy among
the lengths of the sides of the triangle,
enables us to write the three-point function in the following form,
\begin{eqnarray}
\langle \phi_a (t_1,{\vec y_1}) \phi_b (t_2,{\vec y_2}) \phi_c (t_3,{\vec y_3}) \rangle=&&\int d{\pmb \phi}^1 d{\pmb \phi}^2 d{\pmb \phi}^3d{\pmb \phi}_Rd{\pmb \phi}_r~\phi^1_a \phi^2_b \phi^3_c ~\Pi ({\pmb \phi}^1,t_1;{\pmb \phi}_R,t_R) \Pi ({\pmb \phi}^2,t_2;{\pmb \phi}_r,t_r) \nonumber \\
&&\times \Pi ({\pmb \phi}^3,t_3;{\pmb \phi}_r,t_r) \Pi ({\pmb \phi}_r,t_r;{\pmb \phi}_R,t_R) \rho_{\rm eq}({\pmb \phi}_R). 
\end{eqnarray}
Substituting Eq.~(\ref{sol-Pi}), the above expression reduces to
\begin{eqnarray}
\langle \phi_a (t_1,{\vec y_1}) \phi_b (t_2,{\vec y_2}) \phi_b (t_3,{\vec y_3}) \rangle&=&{\cal N}^2 \sum_{\ell,m,n} A_a^{(\ell)}A_b^{(m)}A_c^{(n)} B_{\ell m n} \nonumber \\
&&\times e^{-\Lambda_\ell(t_1-t_R)}e^{-\Lambda_m(t_2-t_r)}
e^{-\Lambda_n (t_3-t_r)} e^{-\Lambda_\ell(t_r-t_R)}, \label{squeezed-3pt}
\end{eqnarray}
where $B_{\ell m n}$ is defined by
\begin{equation}
B_{\ell m n} \equiv \int d{\pmb \phi}~e^{v({\pmb \phi})} \Phi_\ell ({\pmb \phi})\Phi_m ({\pmb \phi})\Phi_n ({\pmb \phi}), \label{def-B}
\end{equation}
and is totally symmetric under the permutation of the indices.

Now let us consider the late time behavior of Eq.~(\ref{squeezed-3pt}).
Since all the eigenvalues satisfy $\Lambda_n \ge 0$,
the leading contribution comes from terms with a particular set of $({\bar \ell},{\bar m},{\bar n})$
having nonvanishing $A_a^{({\bar \ell})}A_b^{({\bar m})}A_c^{({\bar n})} B_{{\bar \ell} {\bar m} {\bar n}}$ 
with the minimum decay rate $e^{-\Lambda_{\bar \ell} t_1}e^{-\Lambda_{\bar m} t_2}e^{-\Lambda_{\bar n} t_3}$.
This means that each integer of $({\bar \ell},{\bar m},{\bar n})$ is determined by 
the lowest value of the eigenvalues with the condition that
$A_a^{({\bar \ell})}A_b^{({\bar m})}A_c^{({\bar n})} B_{{\bar \ell} {\bar m} {\bar n}}$
does not vanish.
Note that this condition does not necessarily fix $({\bar \ell},{\bar m},{\bar n})$
uniquely apart from the trivial permutations and, depending on the interactions among the scalar fields, 
it is possible that there are more than one set of $({\bar \ell},{\bar m},{\bar n})$.
Generally speaking, all the numbers can be different from each other, 
can be partially equal or completely coincide and concrete values of 
$({\bar \ell},{\bar m},{\bar n})$ needs specification of the underlying model.
This may be understood by considering the simplest case, {\it i.e.}, single field case 
in which $B_{111}$ and $B_{112}$ vanish and the lowest contributions are 
either $B_{122} A^{(1)}A^{(2)}A^{(2)}+{\rm perms.}$ 
(when $\Lambda_3 > 2\Lambda_2-\Lambda_1$) or 
$B_{113} A^{(1)}A^{(1)}A^{(3)}+{\rm perms.}$ (when $\Lambda_3 < 2\Lambda_2-\Lambda_1$). 
If the model yields $\Lambda_3 = 2 \Lambda_2-\Lambda_1$ by chance, 
both two contributions decay in time at the same rate and 
none of the two terms can be neglected even at sufficiently late time.
We will come back to the single field case later.

Using the equations for $t_r$ and $t_R$ given by
\begin{equation}
t_r=-\frac{1}{H} \ln \left( \epsilon R H \right),~~~~~t_R=-\frac{1}{H} \ln \left( \epsilon r H \right),
\end{equation}
we find that the three-point functions for sufficiently late time become
\begin{eqnarray}
\langle \phi_a (t_1,{\vec y_1}) \phi_b (t_2,{\vec y_2}) \phi_b (t_3,{\vec y_3}) \rangle \approx &&  
{\cal N}^2 \sum_{\rm min} B_{{\bar \ell} {\bar m} {\bar n}} 
A_a^{({\bar \ell})}A_b^{({\bar m})}A_c^{({\bar n})}  
e^{\Lambda_{\bar \ell}t_1+\Lambda_{\bar m}t_2+\Lambda_{\bar n} t_3} \nonumber \\
&&\times (HR)^{-\frac{2\Lambda_{\bar \ell}}{H}}
(Hr)^{-\frac{\Lambda_{\bar m}+\Lambda_{\bar n}-\Lambda_{\bar \ell}}{H}}, \label{squee}
\end{eqnarray}
where the summation indicated by ``min'' is done for all the possible sets of
$({\bar \ell},{\bar m},{\bar n})$ satisfying the condition mentioned above.
Here, we fixed $\epsilon$ in the same way as the case of two-point functions.
This is the late time three-point functions in the squeezed limit.

This form allows us to write the expressions for the general shape of the
triangle formed by ${\vec x_1},~{\vec x_2}$ and ${\vec x_3}$ by using that
the left-hand side of Eq.~(\ref{squee}) is invariant under the transformation
that preserves de Sitter isometry.
To understand this, notice that we can always move one of the point, 
say ${\vec x_1}$, to a vector having very
long length by performing the transformation (\ref{transform}).
Indeed, if we choose ${\vec b}$ as $-\frac{\vec x_1}{x_1^2}+{\vec \xi}$, 
then we find that the transformed point is ${\vec y_1}={\vec \xi}/\xi^2$, 
whose distance from the origin can be arbitrary large in the limit $\xi \ll 1$. 
By this transformation, time coordinates, which have the same value in the original frame, 
take different values in the new frame.
Thus, equal time correlator for arbitrary configuration of points is related
to the squeezed correlator with different time coordinates for different points
by isometry-preserving transformation.
This is the reason why only the information in the squeezed limit is enough to obtain
the three-point functions for any configuration of points.

The de Sitter isometries for the spatial coordinates at late time $|\eta| \to 0$
become the conformal transformation which serves a base for the conformal field theory \cite{Antoniadis:2011ib}. 
In the language of the conformal field theory \cite{francesco}, 
focusing on any one particular term in Eq.~(\ref{squee}),
the field $\phi_a$ can be interpreted as a primary field of a conformal weight $-\Lambda_a/H$.
Since it is well established how to obtain the general expression of the three-point functions 
of the conformal fields out of the squeezed limit \cite{francesco}, 
we do not expand the detailed discussion here and we only give the final result 
for the three-point functions which is given by
\begin{eqnarray}
\langle \phi_a (t,{\vec x_1}) \phi_b (t,{\vec x_2}) \phi_c (t,{\vec x_3}) \rangle &\approx& 
{\cal N}^2\sum_{\rm min} B_{{\bar \ell} {\bar m} {\bar n}} A_a^{({\bar \ell})}A_b^{({\bar m})}A_c^{({\bar n})} 
{\left( H R_{12} \right)}^{-\frac{\Lambda_{\bar \ell}+\Lambda_{\bar m}-\Lambda_{\bar n}}{H}}  \nonumber \\
&&\times  
{\left( H R_{23} \right)}^{-\frac{\Lambda_{\bar m}+\Lambda_{\bar n}-\Lambda_{\bar \ell}}{H}}
{\left( H R_{31} \right)}^{-\frac{\Lambda_{\bar n}+\Lambda_{\bar \ell}-\Lambda_{\bar m}}{H}}, \label{3pt-final}
\end{eqnarray}
where $R_{ij} \equiv a(t) |{\vec x_i}-{\vec x_j}|$ represents the physical distance
between ${\vec x_i}$ and ${\vec x_j}$.
This equation is one of our primary result.
This shows that all the information regarding the three-point functions can be obtained 
once we know the eigenfunctions and eigenvalues of
the $N$ dimensional Schr\"odinger equation (\ref{Schrodinger}).
As a consistency check, we can verify that squeezed limit of Eq.~(\ref{3pt-final}) 
($R_{12}=R_{13}=R,~R_{23}=r$) gives back Eq.~(\ref{squee}).
Also, we can implement the similar derivation of the three-point functions for
the equilateral case in which $|{\vec y_1}-{\vec y_2}|=|{\vec y_2}-{\vec y_3}|=|{\vec y_3}-{\vec y_1}|$.
This case also allows the evaluation of the three-point function without resorting to the
direct computation of Eq.~(\ref{evo-rho3}).
It can be verified that the result coincides with the equilateral case of Eq.~(\ref{3pt-final}).

Now there are several points to be remarked.
First, as mentioned earlier, $({\bar \ell},{\bar m},{\bar n})$ is given by
the condition that it is a set of integers as small as possible with nonvanishing
$A_a^{({\bar \ell})}A_b^{({\bar m})}A_c^{({\bar n})} B_{{\bar \ell} {\bar m} {\bar n}}$.
In principle, this integer set can vary for different choice of fields $(a,b,c)$.
This suggests that only a knowledge of correlators of the three product of the same field
({\it i.e.}, $\langle \phi_a \phi_a \phi_a \rangle$ etc.) is not enough to know the scaling behavior of the
three-point function of the three different fields.
Furthermore, if at least one of the field appearing in the correlator is different from others,
it can happen that the correlator cannot be given by a single term with a power-law form.
Instead the correlator becomes a sum of up to six terms each of which exhibits 
the different power-law behavior.
Notice that all of those terms are not necessarily nonvanishing and it is also possible that
only some of them remain nonzero.

Secondly, the result (\ref{3pt-final}) is obtained without the use of the perturbative expansion
in terms of the strength of the interactions among fields.
In the standard approach, the three-point functions (and higher order functions as well) 
are calculated by using the so-called
in-in formalism which usually uses perturbative expansion and truncation of the calculations
at some order to yield an analytic expression.
Although this approach is completely justified as long as the higher-order terms contribute 
much less to the final result,
it is known that some particular model (for example, massless scalar field with a quartic
self-interaction) requires nonperturbative treatment to obtain the reliable correlators.
In more general terms, the perturbative approach fails when the system does possess the 
de Sitter invariant state only if the interactions are present.
In such a case, the standard perturbative approach needs some care, if not impossible,
to get the correct result. 
On the other hand, our result (\ref{3pt-final}) is derived by the stochastic approach.
As is well known, the stochastic approach includes the nonperturbative effects coming 
from the long wavelength modes.
Except for some simple models, numerical computations are required to solve Eq.~(\ref{Schrodinger})
in order to obtain the eigenvalues $\Lambda_n$.
But whichever computation method is used, the obtained $\Lambda_n$ contains the nonperturbative 
effect. 
This is also true for the amplitude of the correlator, {\it i.e.},  
$A_a^{({\bar \ell})}A_b^{({\bar m})}A_c^{({\bar n})} B_{{\bar \ell} {\bar m} {\bar n}}$.
The point is that everything is reasonably calculable in the stochastic approach
while it is hard to take into account the higher-order or nonperturbative effects
in the standard perturbative expansion of the in-in formalism.
Therefore, our result will be quite useful when the nonperturbative effect is crucial
to get the correct correlators.

\subsection{Four-point functions}
Contrary to the case of the three-point function where consideration in the squeezed 
limit is enough to get the correlator for the general triangle,
this is not the case for the four-point function.
The reason behind this is that the transformation specified by the vector ${\vec b}$ is
not sufficient to convert any quadrangle to the one having hierarchy among all the sides
of the quadrangle.
Thus we need to solve the evolution equation for the probability distribution function
for four variables even to obtain the late time behaviour of the correlator.
Yet, it would be interesting to see to what extent we can restrict the form of the 
four-point function in analytic way.

\begin{figure}[t]
  \begin{center}{
    \includegraphics[scale=0.7]{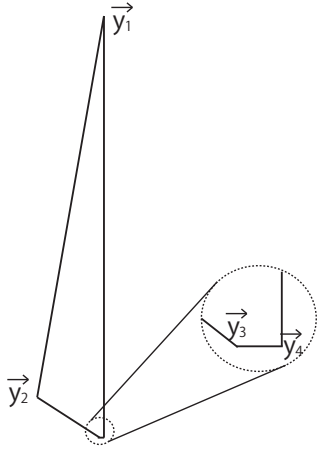}
    }
  \end{center}
  \caption{Double squeezed quadrangle for which
$R_{14} \simeq R_{12} \gg R_{23} \simeq R_{24} \gg R_{34}$,
where $R_{ij}$ denotes the physical distance between ${\vec y}_i$ and ${\vec y}_j$.}
 \label{fig1}
\end{figure}

Let us first consider the double squeezed quadrangle for which
$R_{14} \simeq R_{12} \gg R_{23} \gg R_{34}$,
where $R_{ij}$ denotes the distance between ${\vec y}_i$ and ${\vec y}_j$ (see Fig.~\ref{fig1}).
For this quadrangle, using the similar reasoning we made in the case
of the three-point function, the four-point function can be written as
\begin{eqnarray}
\left\langle \prod_{i=1}^4 \phi_{a_i} (t_i,{\vec y_i}) \right\rangle  =&&{\cal N}^2 \sum_{m,n,p,q,s}B_{mns} B_{spq}A^{(m)}_{a_1} A^{(n)}_{a_2} A^{(p)}_{a_3} A^{(q)}_{a_4} e^{-\Lambda_m (t_1-t_X)-\Lambda_n (t_2-t_Y)}\nonumber \\
&&\times e^{-\Lambda_p (t_3-t_Z)-\Lambda_q (t_4-t_Z)-\Lambda_s (t_Z-t_Y)-\Lambda_m (t_Y-t_X)},
\end{eqnarray}
where $X =R_{14},~Y=R_{24}$ and $Z=R_{34}$ ($X \gg Y \gg Z$).
Interestingly, the four-point function in this limit is written solely in terms of 
the quantities characterizing the three-point functions.
At sufficiently late time, this reduces to
\begin{eqnarray}
\left\langle \prod_{i=1}^4 \phi_{a_i} (t_i,{\vec y_i}) \right\rangle  &&={\cal  N}^2 \sum_{\rm min}B_{{\bar m}{\bar n}{\bar s}} B_{{\bar s}{\bar p}{\bar q}}A^{({\bar m})}_{a_1} A^{({\bar n})}_{a_2} A^{({\bar p})}_{a_3} A^{({\bar q})}_{a_4} e^{-\Lambda_{\bar m} t_1-\Lambda_{\bar n} t_2}\nonumber \\
&&\times e^{-\Lambda_{\bar p} t_3-\Lambda_{\bar q} t_4} {\left( HX \right)}^{-\frac{2\Lambda_{\bar n}}{H}} 
{\left( HY \right)}^{-\frac{\Lambda_{\bar m}+\Lambda_{\bar s}-\Lambda_{\bar n}}{H}}
{\left( HZ \right)}^{-\frac{\Lambda_{\bar p}+\Lambda_{\bar q}-\Lambda_{\bar s}}{H}}, \label{4pt-doublesqueeze}
\end{eqnarray}
where the meaning of the ``min'' in the summation is the same as the case for the 
three-point function.
Now, the expression of the right-hand side of the above equation must be the
double squeezed limit of the correlator for the general quadrangle.
Using again the procedure of restricting the form of the four-point function 
in the conformal field theory \cite{francesco}, 
we find that the four-point function (\ref{4pt-doublesqueeze}) for the general quadrangle becomes
\begin{eqnarray}
\left\langle \prod_{i=1}^4 \phi_{a_i} (t,{\vec x_i}) \right\rangle &&=
{\cal N}^2 \sum_{\rm min}f_{a_1a_2a_3a_4}^{{\bar m}{\bar n}{\bar p}{\bar q}} \left( \frac{R_{12}R_{34}}{R_{13}R_{24}},~\frac{R_{12}R_{34}}{R_{23}R_{14}} \right) \prod_{i<j} {\left( HR_{ij} \right)}^{\frac{\Delta}{3}-\Delta_i-\Delta_j}, \label{4pt-general}
\end{eqnarray}
where $\Delta_1=\frac{\Lambda_{\bar m}}{H},~\Delta_2=\frac{\Lambda_{\bar n}}{H}$ 
etc. and $\Delta=\Delta_1+\Delta_2+\Delta_3+\Delta_4$.
The function $f_{a_1a_2a_3a_4}^{{\bar m}{\bar n}{\bar p}{\bar q}}$ depending on the two
arguments that are invariant under the de Sitter isometric transformation 
(only for late time or on superhorizon scales) is completely arbitrary function 
at the level of the symmetry argument.
Some more nontrivial information is needed to (even partially) fix the form of it.

\begin{figure}[t]
  \begin{center}{
    \includegraphics[scale=0.7]{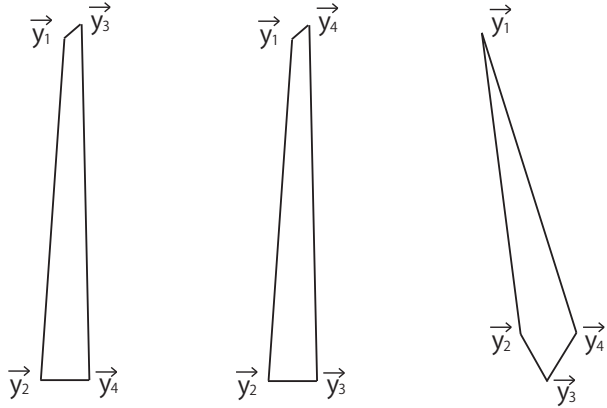}
    }
  \end{center}
  \caption{Left quadrangle having $R_{12} \approx R_{34} \gg R_{13} \approx R_{24}$ yields Eq.~(\ref{case1}).
  Middle quadrangle having $R_{12} \approx R_{34} \gg R_{14} \approx R_{23}$ yields Eq.~(\ref{case2}).
  Right quadrangle having $R_{12} \approx R_{13} \approx R_{14} \gg R_{23} = R_{34} = R_{24}$ yields
  Eq.~(\ref{case3}).}
 \label{fig2}
\end{figure}

Although we cannot find analytic form of $f_{a_1a_2a_3a_4}^{{\bar m}{\bar n}{\bar p}{\bar q}}$
for the whole domain of the arguments,
we can derive its asymptotic behavior or a value at specific point for some limiting cases of the arguments 
by considering the corresponding squeezed shape of the quadrangle.
For instance, comparison between Eqs.~(\ref{4pt-doublesqueeze}) and (\ref{4pt-general}) leads to the 
following asymptotic form of $f_{a_1a_2a_3a_4}^{{\bar m}{\bar n}{\bar p}{\bar q}} (z,z)$ when
$z \ll 1$ as
\begin{equation}
f_{a_1a_2a_3a_4}^{{\bar m}{\bar n}{\bar p}{\bar q}}(z,z) \simeq B_{{\bar m}{\bar n}{\bar s}} B_{{\bar s}{\bar p}{\bar q}}A^{({\bar m})}_{a_1} A^{({\bar n})}_{a_2} A^{({\bar p})}_{a_3} A^{({\bar q})}_{a_4} z^{\frac{\Lambda_{\bar s}}{H}-\frac{\Lambda_{\bar m}+\Lambda_{\bar n}+\Lambda_{\bar p}+\Lambda_{\bar q}}{3H}},~~~~~{\rm for}~~ z\ll 1. \label{trib}
\end{equation}
In a similar way, 
consideration of the other quadrangles with different shapes shown in Fig.~\ref{fig2}
leads to the following expression;
\begin{eqnarray}
&&f_{a_1a_2a_3a_4}^{{\bar m}{\bar n}{\bar p}{\bar q}}(z,1) \simeq B_{{\bar m}{\bar n}{\bar s}} B_{{\bar s}{\bar p}{\bar q}}A^{({\bar m})}_{a_1} A^{({\bar n})}_{a_2} A^{({\bar p})}_{a_3} A^{({\bar q})}_{a_4} z^{-\frac{\Lambda_{\bar s}}{H}+\frac{\Lambda_{\bar m}+\Lambda_{\bar n}+\Lambda_{\bar p}+\Lambda_{\bar q}}{3H}},~~~~~{\rm for}~~ z\gg 1, \label{case1}\\
&&f_{a_1a_2a_3a_4}^{{\bar m}{\bar n}{\bar p}{\bar q}}(1,z) \simeq B_{{\bar m}{\bar q}{\bar s}} B_{{\bar s}{\bar n}{\bar q}}A^{({\bar m})}_{a_1} A^{({\bar n})}_{a_2} A^{({\bar p})}_{a_3} A^{({\bar q})}_{a_4} z^{-\frac{\Lambda_{\bar s}}{H}+\frac{\Lambda_{\bar m}+\Lambda_{\bar n}+\Lambda_{\bar p}+\Lambda_{\bar q}}{3H}},~~~~~{\rm for}~~ z\gg 1, \label{case2}\\
&&f_{a_1a_2a_3a_4}^{{\bar m}{\bar n}{\bar p}{\bar q}}(1,1) \simeq T_{{\bar m}{\bar n}{\bar p}{\bar q}}A^{({\bar m})}_{a_1} A^{({\bar n})}_{a_2} A^{({\bar p})}_{a_3} A^{({\bar q})}_{a_4}, \label{case3}
\end{eqnarray}
In Eq.~(\ref{case3}), we have introduced $T_{{\bar m}{\bar n}{\bar p}{\bar q}}$ defined by
\begin{equation}
T_{{\bar m}{\bar n}{\bar p}{\bar q}}=\int d{\pmb \phi}~e^{2v({\pmb \phi})} \Phi_m ({\pmb \phi})\Phi_n ({\pmb \phi}) \Phi_p ({\pmb \phi})\Phi_q ({\pmb \phi}),
\end{equation}
which is a new quantity to characterize the four-point function.
These expressions for the several limiting cases indicate that four-point function
having the double squeezed quadrangle is determined by $B_{mns}$, {\it i.e.},
related to (square of) the three-point function while the correlator
having the single squeezed one is determined by $T_{mnpq}$ which has nothing to
do with the three-point function.
Finally, let us remark that Eqs.~\eqref{trib} -- \eqref{case3} cover all the possible limiting cases which we can obtain. One can show that we cannot make such a quadrangle which provides $f_{a_1a_2a_3a_4}^{{\bar m}{\bar n}{\bar p}{\bar q}}(z,w)$ with $z=1,~w\ll 1$, or $z\ll 1,~w=1$ or $z,w\gg 1$.

\section{Discussion and summary}
In this paper, we have developed a general formalism to calculate three- and
four-point functions among test light scalar fields in de Sitter space in the framework
of the stochastic formalism.
Stochastic approach treats the coarse grained fields as classical variables that are
affected by the random Gaussian noises reflecting the effect that the short wavelength 
modes enter the long wavelength modes due to the accelerated expansion.
We investigated behaviors of those correlators at sufficiently late time when the fields at different
spatial points evolve independently and derived general formulae of the correlators.
In the stochastic approach, one generally needs to solve the Focker-Planck equation
to evaluate the time evolution of the correlator since its solution (probability density) 
appears in the definition of the correlator.
This program (with a reasonable approximation) works for the two-point functions and the
result is given by Eq.~(\ref{two-point-2}).
This result clearly shows that the amplitude of the two-point functions is determined by
$A^{(n)}_a$ defined by Eq.~(\ref{def-A}) which represents the expectation value of 
$\phi_a$ for the $n$-th state $e^{-v({\pmb \phi})} \Phi_n ({\pmb \phi})$
corresponding to the eigenvalue $\Lambda_n$ which determines the scaling index. 
It is important to notice that by construction the stochastic formalism incorporates 
the dynamics of the long wavelength modes without expanding the basic equations in terms 
of the amplitude of the coarse grained fields or the strength of the interactions.
Therefore, once $A^{(n)}_a$ and $\Lambda_n$ and hence the two-point functions
are obtained, they are the results that do not rely on the perturbative expansion
and automatically involve the nonperturbative effects.
Except for some simple cases, Eq.~(\ref{Schrodinger}) does not allow the analytic solution
and numerical computation is generally required to obtain the concrete values of $A^{(n)}_a$
and $\Lambda_n$.
However, solving Eq.~(\ref{Schrodinger}) is a mathematically well-defined problem
and, in principle, it can be solved (especially for the case with a small number of fields) 
without any principal difficulty.
This plausible feature of the stochastic formalism is in striking contrast to the standard calculation of
the correlation function that usually uses perturbative expansion in terms of the
strength of the interactions and truncates the expansion at some order.
In most cases, the truncation is done at the lowest order that yields nonvanishing contribution
and extension to the higher order entails very messy expression that requires careful 
consideration to obtain physically meaningful results.

Direct manipulation of the evolution equation for the probability density becomes very difficult
for the case of the three-point functions.
Therefore, in this paper, we adopted another approach which works as long as we are only 
interested in the late time behavior of the correlators.
The basic idea is to utilize the de Sitter isometries and to consider only the squeezed case 
in which one of the three spatial points is separated far way than the others.
This limiting case allows us to build up the analytic expressions of the correlators at late time. 
We then converted the correlators to the ones for the arbitrary configuration of points by using the fact
that the value of the correlator itself remains the same under the isometric transformation
for the de Sitter invariant state.
Among the ten de Sitter isometries, the three reduces to the special conformal transformation
for the spatial transformation.
It is this transformation that enables us to transform any triangle into the arbitrarily squeezed one
and to obtain the analytic form of the (only late time) three-point function for 
any configuration of points which is given by Eq.~(\ref{3pt-final}).

Now it would be interesting to consider the consequences of Eq.~(\ref{3pt-final})
focusing on the single field case for simplicity.
Even in this case, we find nontrivial and interesting properties of the three-point function.
Noting that we can always make $A^{(0)}$ be zero by suitably redefining the field,
the leading term that remains at late time is given by
$B_{111} {(A^{(1)})}^3$ provided neither $B_{111}$ nor $A^{(1)}$ vanishes.
Since this combination $(1,1,1)$ provides the lowest decaying rate of the correlator,
this is the only leading term that survives at late time.
Therefore, Eq.~(\ref{3pt-final}) in this case becomes
\begin{equation}
\langle \phi (t,{\vec x_1}) \phi (t,{\vec x_2}) \phi (t,{\vec x_3}) \rangle = {\cal N}^2
B_{111} {(A^{(1)})}^3 {\left(HR_{12} \right)}^{-\frac{\Lambda_1}{H}}{\left(HR_{23} \right)}^{-\frac{\Lambda_1}{H}}
{\left(HR_{31} \right)}^{-\frac{\Lambda_1}{H}}. \label{3pt-single}
\end{equation}
In this case, two-point function at late time becomes
\begin{equation}
\langle \phi (t,{\vec x_1}) \phi (t,{\vec x_2}) \rangle ={\cal N}^2 
{(A^{(1)})}^2 {\left(HR_{12} \right)}^{-\frac{2\Lambda_1}{H}}. \label{2pt-single}
\end{equation}
Scaling behavior of Eqs.~(\ref{3pt-single}) and (\ref{2pt-single}) coincide with 
the one given in \cite{Antoniadis:1996dj,Antoniadis:2011ib,Creminelli:2011mw,Kehagias:2012pd}.
In these references, the scaling behavior was derived by combining the de Sitter isometries like we have done 
in this paper and the assumption that $\phi$ (not at the
level of the correlators) at sufficiently late time scales as
$\phi \sim {(-\eta)}^\Delta$ and this scaling directly enters the scaling 
of the correlator, for instance, 
$\langle \phi(\eta_1) \phi (\eta_2) \rangle \sim {(-\eta_1)}^\Delta {(-\eta_2)}^\Delta$.
On the other hand, as we have shown, the stochastic formalism can provide 
a necessary and sufficient condition in order for the above
naive assumption to hold, which is given by $A^{(1)} B_{111} \neq 0$.
In addition to this, the stochastic formalism also gives amplitudes of the 
correlators in terms of the calculable quantities.

What happens if $B_{111}$ accidentally vanishes, which is possible for some models?
In this case, the leading contribution to the three-point function comes from
a term $B_{112} {(A^{(1)})}^2 A^{(2)}$ unless it vanishes
\footnote{If the potential has a reflection symmetry $V(\phi)=V(-\phi)$,
$\Phi_1(\phi)$ for the bound state is an odd function,
so that $B_{111}$ in Eq.~(\ref{def-B}) vanishes.
In this case, however,
$B_{112}$ should also vanish because of the same symmetry.}.
Then, the two-point function is still given by Eq.~(\ref{2pt-single}),
but the three-point function becomes
\begin{equation}
\langle \phi (t,{\vec x_1}) \phi (t,{\vec x_2}) \phi (t,{\vec x_3}) \rangle = {\cal N}^2
B_{112} {(A^{(1)})}^2 A^{(2)}
\bigg[ {\left(HR_{12} \right)}^{-\frac{2\Lambda_1-\Lambda_2}{H}}
{\left(HR_{23} \right)}^{-\frac{\Lambda_2}{H}} {\left(HR_{31} \right)}^{-\frac{\Lambda_2}{H}}+2~{\rm perms.} \bigg].
\label{b112}
\end{equation}
This is very different from Eq.~(\ref{3pt-single}) on two counts;
it does not obey the single power-law, and it is not given by a single term
but by three terms that are mutually related by permutations.
As far as we know, this type of three-point functions has been overlooked in literature.
If $B_{112} {(A^{(1)})}^2 A^{(2)}$ vanishes too, then we need to consider 
$B_{122}A^{(1)}{(A^{(2)})}^2 $ or $B_{113}A^{(1)}{(A^{(2)})}^2$, whichever yields
the lower decaying rate.
If both of these have the same decaying rate, we must keep both of them in the correlator.
Obviously, any of these leads to the multi-scaling expressions of the
three-point function like Eq.~(\ref{b112}). 
These examples suggest that knowledge of the scaling behavior of the two-point
function is not necessarily enough to know the scaling behavior of the
three-point function.
In principle, three-point function can exhibit more complicated structure than the
native expectation.

The single field with $B_{111}\neq 0$ gives the following form of the four-point function,
\begin{equation}
\left\langle \prod_{i=1} \phi (t,{\vec x_i}) \right\rangle =
{\cal N}^2 f^{1111} \left( \frac{R_{12}R_{34}}{R_{13}R_{24}},~\frac{R_{12}R_{34}}{R_{23}R_{14}} \right) \prod_{i<j} {\left( HR_{ij} \right)}^{-\frac{2\Lambda_1}{3H}}.
\end{equation}
This expression is the same as the one given in \cite{Creminelli:2011mw}.
The function $f^{1111}$ for some limiting cases are given in Eqs.~(\ref{trib})-(\ref{case3}).
For instance, we find
\begin{equation}
f^{1111}(z,z) =B_{111}^2 {(A^{(1)})}^4 z^{-\frac{\Lambda_1}{3H}},~~~~~{\rm for}~z\ll 1. \label{single-4pt}
\end{equation}
Generally speaking, $B_{111}=0$ does not imply the vanishing of $f^{1111}$
but only changes the form of $f^{1111}$.
Assuming that $B_{112}$ does not vanish, $f^{1111}$ corresponding to 
Eq.~(\ref{single-4pt}) now becomes
\begin{equation}
f^{1111}(z,z) =B_{112}^2 {(A^{(1)})}^4 z^{\frac{\Lambda_2}{H}-\frac{4\Lambda_1}{3H}},~~~~~{\rm for}~z\ll 1,
\end{equation}
which contains information of the second excited state $\Phi_2$.
As it should be, the power of $z$ is higher by $(\Lambda_2-\Lambda_1)/H$
than Eq.~(\ref{single-4pt}).\\

\noindent {\bf Acknowledgments:} 
We would like to thank A.~Riotto and A.~Kehagias for explaining the 
applicability of the results of \cite{Kehagias:2012pd}.
This work was supported by JSPS Research Fellowships for Young Scientists (HM), 
Grant-in-Aid for JSPS Fellows No.~1008477 (TS), 
JSPS  Grant-in-Aid for Scientific Research No.\ 23340058 (JY), 
and the Grant-in-Aid for Scientific Research on Innovative Areas No.\ 21111006 (JY).

\end{document}